\documentclass[showpacs,amsmath,amssymb]{revtex4}

\usepackage{graphicx}
\usepackage{color}

\begin{document}

\title{Two-dimensional electron gas at the interface of \\
the ferroelectric-antiferromagnetic heterostructure
Ba$_{0.8}$Sr$_{0.2}$TiO$_3$/LaMnO$_3$}
\author{D.\,P.\,Pavlov$^{+}$, I.\,I.\,Piyanzina$^{+,*}$,
V.\,M.\,Mukhortov$^{\#}$, A.\,M.\,Balbashov$^{\&}$,
D.\,A.\,Tayurskii$^{*}$, I.\,A.\,Garifullin$^{+}$,  \\
R.\,F.\,Mamin$^{+,*}$\/\thanks{e-mail:mamin@kfti.knc.ru}}

\affiliation{$^{+}$ E. K. Zavoisky Physical-Technical Institute,
420029 Kazan, Russia,
\\~\\
$^{*}$  Institute of Physics, Kazan Federal University, 420008 Kazan, Russia,
\\~\\
$^{\#}$  Southern Scientific Center of Russian Academy of Sciences, 344006 Rostov-on-Don, Russia,
\\~\\
$^{\&}$ Moscow Power Engineering Institute, 111250 Moscow, Russia }

\begin{abstract}
The temperature dependence of the electrical resistivity of the
heterostructures consisting of single crystalline LaMnO$_3$ samples
with different crystallographic orientations covered by the
epitaxial ferroelectric Ba$_{0.8}$Sr$_{0.2}$TiO$_3$ film has been
studied. Results obtained for the heterostructure have been compared
with the electrical resistivity of the single crystalline LaMnO$_3$
without the film. It was found that for the samples with the films
where the polarization axis is perpendicular to the crystal surface
the electrical resistivity strongly decreases, and at the
temperature below ~160\,K undergoes the insulator-metal transition.
$ Ab-initio $ calculations were also performed for the structural
and electronic properties of the BaTiO$_3$/LaMnO$_3$
heterostructure. Transition to the 2D electron gas at the interface
is shown.
\end{abstract}

\maketitle

The high-mobility electron gas has been observed the first time in
2004 \cite{S1} at the interface between LaAlO$_3$ (LAO) and
SrTiO$_3$ (STO). Later on similar heterostructures consisting of a
nonmagnetic and insulating oxides have been extensively studied. In
particular, it was found that nanometer-thick metallic phase
(two-dimensional electron gas - 2DEG) is formed in the STO layers at
the LAO/STO interface when the number of LAO overlayers exceeds
three~\cite{S2}. Below 300 mK~\cite{S3} the system passes into the
superconducting state. The density of the current carriers in such
heterostructure reaches the value of 3$\times$10$ ^{13}$\,cm$^{-2}$.
In addition, the ferromagnetic order has also been found in the
system~\cite{S4}. 2DEG have been later found in other non-magnetic
dielectrics, e.g., in KTaO$_3$/SrTiO$_3$~\cite{S5} and
CaZrO$_3$/SrTiO$_3$~\cite{S6}. Also, 2DEG has been found at the
interface between Mott-insulators and a magnetic layer, in
particular, with ferromagnetic GdTiO$_3$~\cite{S7}, and also at the
interface with antiferromagnetic  SmTiO$_3$~\cite{S8},
LaTiO$_3$~\cite{S9}, with maximum possible electron density of
3$\times $10$^{14}$\,cm$^{-2}$. Later on~\cite{S11},  formation of
2DEG has been demonstrated in NdAlO$_3$/SrTiO$_3$,
PrAlO$_3$/SrTiO$_3$, NdGaO$_3$/SrTiO$_3$, and also in
LaGaO$_3$/SrTiO$_3$~\cite{S10} heterostructure.

It is supposed that the appearance of conductivity is coupled with
the structural and, consequently, electronic reconstructions.
However,  so far there is no unambiguous explanation and the theory
which can be able to explain all phenomena found in those systems.
One of the most important feature related to the 2DEG formation is
the local polarity of the (LaO)$^{+1}$ and (AlO$_2)^{-1}$ layers
inside the LAO slab.
In this paper in order to investigate the correlations between
structural distortions, electronic reconstruction  and polarity, we
have chosen the BaTiO$_3$/LaMnO$_3$ (BTO/LMO) heterostructure, where
all layers BaO and TiO$_2$ in simple electronic limit are
"electrically neutral", but there is a ferroelectric polarization
due to the Ti$^{+4}$ atoms displacements out of octahedron center in
the BTO. The direction of such a polarization can be switched by an
external electric field, what is impossible to be done in the LAO
slab, because an external influence cannot change the sequence of
(LaO)$ ^{1+} $ and AlO$ _{2} ^{1-}$ layers. Moreover it is very
important that {\bf in our case there is no need to make a very
high-quality ferroelectric-antiferromagnetic boundary, because the
polarization arises in the volume of the ferroelectric (This differs
from the case of LAO where, for the appearance of polarization on
the interface, it is necessary to obey strictly the sequence of
(LaO)$^{+1}$ and (AlO$_2)^{-1}$ layers)}. Besides, the BTO/LMO
system is attract the interest because it contains antiferromagnetic
LaMnO$_3$, which can be transferred to ferromagnetic state by
increasing the concentration of the free carriers
\cite{deGennes,Dog} (the reason for that is an increasing the
indirect ferromagnetic exchange). That can be realized by doping of
the LMO \cite{deGennes,Dog}. It can be further expected that
increasing the free charge carriers can lead to the local
ferromagnetic order and magnetoresistivity~\cite{Dog} in a system
with 2DEG. Therefore, there is an opportunity to switch both
conductivity by an electric field (trigger effect), and the magnetic
order (magnetoelectric effect) in the heterostructures similar to
BTO/LMO. We present the results on  of calculation of the structural
and electronic properties calculations for the BaTiO$_3$/LaMnO$_3$
(BTO/LMO) heterostructure, as well as experimental study  of the
resistivity of the Ba$_{0.8}$Sr$_{0.2}$TiO$_3$/LaMnO$_3$ (BSTO/LMO)
heterostructure with 350\,nm thickn BSTO layer.

\par
For the experimental study we used the single crystalline LMO. On
their modified surface the epitaxial films of the BSTO were
deposited using the sputtering technique. Our sample choice is based
on the following facts: (1) preparation technique for the single
components was previously well developed and tested~\cite{S23, S21,
S22}; (2) properties of Вa$ _{0.8} $S$ _{0.2} $TO$ _{3} $ and BTO
are not much different; (3) transition temperature to the
ferroelectric state $T_c$ is relativity high~\cite{S21}:
$T_c\sim$540\,K for 300\,nm-thick films~\cite{S21}. Properties of
2DEG slightly change starting from the certain thickness of the part
of the system which demonstrates polar properties indicated by our
modeling. That is why we used 350\,nm-thick films.
In the present investigations it was important to find conducting
state and to fix it reliably whereas the influence of the thickness
of the \textbf{}ferroelectric component on the properties of 2DEG is
the aim of our further investigations.
The thin film of Ba$_{0.8}$Sr$_{0.2}$TiO$_3$ (BSTO)  was sputtered
on the top of the single crystalline LaMnO$_3$  samples. The
resistivity measurements were performed by a four-point probe
method. In parallel we started the first-principles calculations in
order to determine structural and electronic properties of the
BTO/LMO heterostructure.

\par
First of all, we present the results of our calculations of the
properties of the BTO/LMO heterostructure. We used density
functional theory (DFT) as a main method~\cite{S14}. Exchange and
correlational effects were accounted by generalized gradient
approximation (GGA)~\cite{S17}. Kohn-Sham equations were solved
using the plane-wave basis set (PAW)~\cite{S18}, realized within the
VASP code~\cite{S15}, implemented into the MedeA
software~\cite{S19}. The cut-off energy were chosen to be 400\,eV,
The force tolerance was 0.05~eV/\AA\ and the energy tolerance for
the self-consistency loop was $ 10^{-5} $~eV. The Brillouin zones
were sampled including $ 5 \times 5 \times 1 $ $ {\bf k} $-points.
Since there is a strong correlation between $ d $ и  $ f $-electrons
in our system the GGA+$ U $ correction were added to our
computational scheme \cite{S16}. The $ U $ parameter were added to
La\,4$ f $, Ti\,3$ d $ и Mn\,4$ d $ ($ U $=8\,eV, 2\,eV и 4\,eV,
respectively). All calculations were performed taking into account
magnetic nature of the material.

%
\begin{figure}
  \centering{\includegraphics[angle=-90,width=3.5cm]{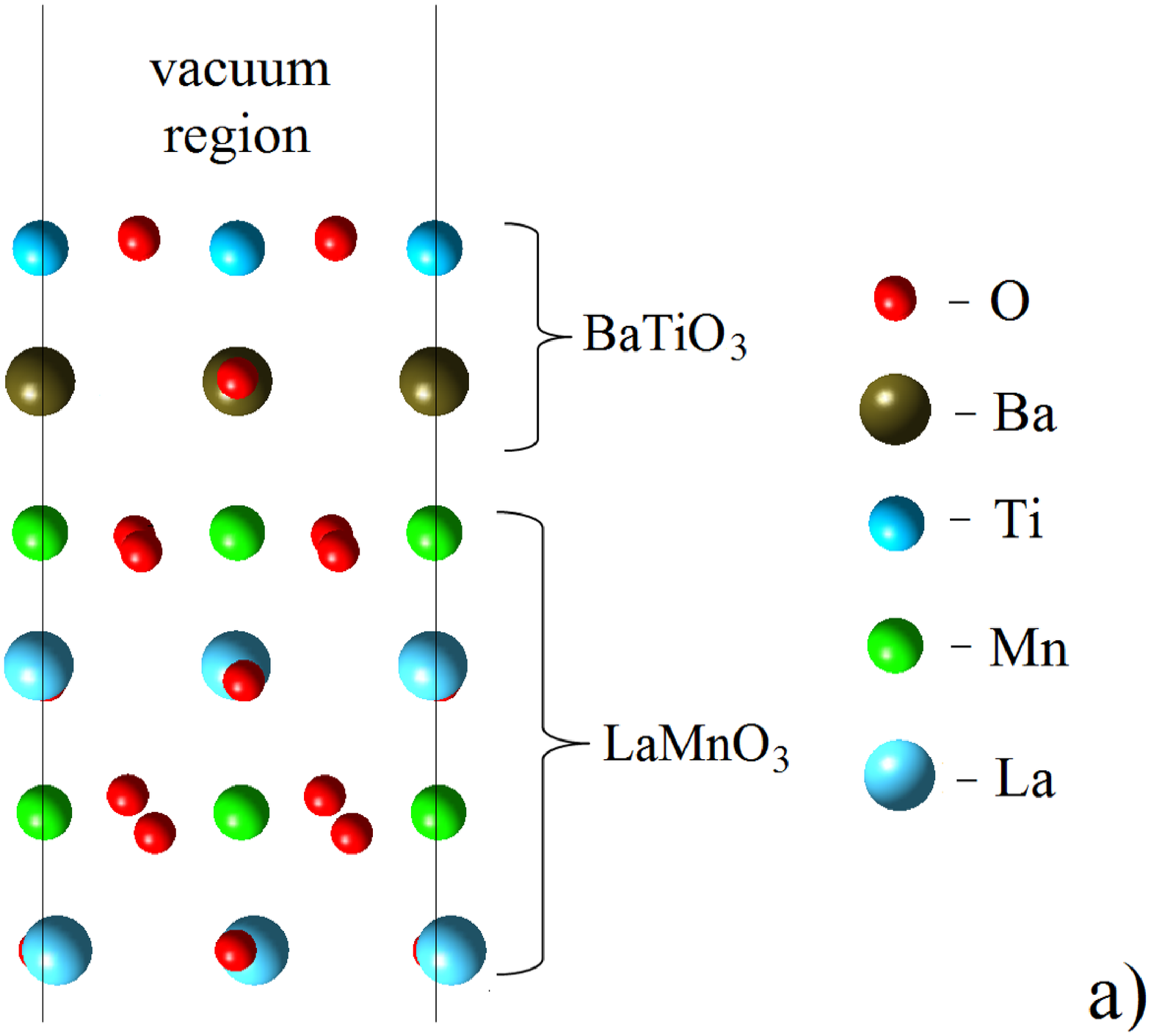}}
 {\includegraphics[angle=0,width=4.52cm]{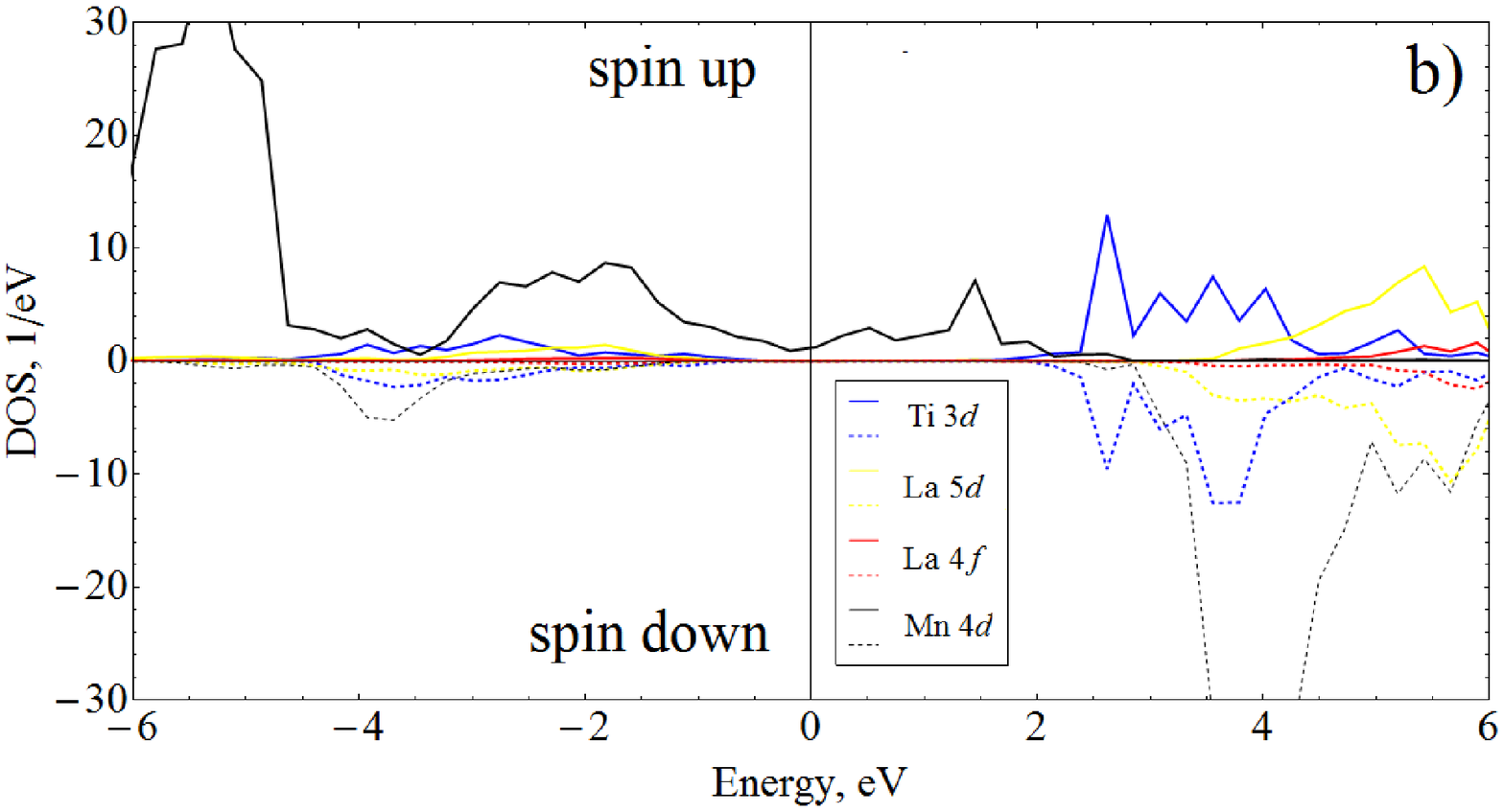}}
   \caption{ Half of the unit cell of the LMO/BTO heterostructure with a  BaO layer at the interface (a).
   Corresponding density of states (b)}
    \label{structure}
\end{figure}

\par
In Fig.\,1a half of the unit cell of the studied system LMO/BTO with
BaO interface layer is presented whereas the second part is a mirror
copy with respect to the central LaO layer. Cell parameters  for the
bulk LMO with orthorhombic structure obtained after optimization
equal to $a$=5.650\,\AA, $b$=5.616\,\AA, $c$=7.935\,\AA\
(experimental values: $a$=5.537\,\AA, $b$=5.749\,\AA,
$c$=7.665\,\AA). For modeling the heterostructure the LMO central
slab  was enlarged by 1.5 times and bounded  by a varing number of
BaTiO$_3$ layers with interface BaO or TiO$_2$ layers on both sides.
Such a unit cell guarantee the absence of the dipole moment and
additional polarity which might arise due to non-symmetric
structure. In order to avoid interaction of the surfaces and slabs
with their periodic images, a 20~\AA\ vacuum region was added. In
plane $a$ и $b$ cell parameters were fixed, whereas atom positions
were allowed to relax during the optimization procedure.

After the optimization of both interface types  (first corresponds
to BaO interface layer, second -- to the TiO$ _{2} $ layer) it was
found that in the first case, the total energy of the system is
lower, what means that the structure is more stable. That is why all
further reasoning will be presented to the most stable
configuration.
It is seen from Fig.~1a that in the near-surface TiO$ _{2} $ layer
the Ti atoms move out of the oxygen planes by  $a  \approx
$0.15\,\AA. That leads to a dipole moment induction towards the
interface. Calculations involving higher number of the BTO layers
are required to get a full picture of structural distortions, what
will be done in our further publications.

In order to determine the electronic properties of the studied
structure the density of states (DOS) spectrum has been calculated
taking into account magnetic features of LMO. Fig.\,1b shows the
atom-resolved DOS for Mn, La and Ti. It is seen that already with
one BTO layer the band gap is closed. Mn atoms orbitals cross the
Fermi-level. Besides, the total magnetic moment induction takes
place which is mainly corresponds to Mn atoms forms.

Measurements were performed for three types of samples: (1) sample
N1 is a heterostructure based on the single crystalline LMO with a
BSTO film on the top of it ($c$ axis of the LMO is parallel to the
deposition plane); (2) sample N2 is a heterostructure based on the
single crystalline  LMO with a BSTO film on top of it ($c$ axis is
perpendicular to the deposition plane); (3) samples N01 and N02 are
the single crystalline LMO  without films with polarization axis as
in the cases of N1 and N2, respectively.
X-ray measurements has shown that: 1) for the N1 sample the
$\textbf{c}$ axis (along which the spontaneous polarization is
detected) is parallel to the film plane; (2) for the N2 samples
$\textbf{c}$ axis is perpendicular to the film plane. The reason for
that is the lattice mismatch: in the first case the LMO substrate is
stretchy with respect to the film, whereas in the second case --
compressive~\cite{S22}. In both cases this is determined by the
difference in unit cell parameters of the single crystal and film.
The electrical resistivity measurements $\rho (T)$ were performed
using the standard four-point method for N01 and N02 samples (the
current is perpendicular to the  $c$ axis of the LMO single
crystal). The results for N2 sample are presented in Fig.~2. The
electrical resistivity shows an activation behavior which is typical
for the semiconducting type of conductivity and corresponds to the
behavior of the electrical resistivity of the LMO, measured for the
similar samples~\cite{Dog, S23}.
For modeling the behavior of the specific electrical resistivity we
used the formula:  $\rho (T)=\rho_{0}\cdot $exp$(T_{a}/T)$ (is shown
in Fig.\,2 by red dashed line for the surface orientation
perpendicular to the $c$ axis). Using it we got the following values
of constants: $\rho_{0}=4.64\cdot $10$^{-3}$\,$ \Omega
$$\cdot $cm, $T_{a}=2938\,K$.

\begin{figure}
 \includegraphics[angle=-90,width=11.05cm]{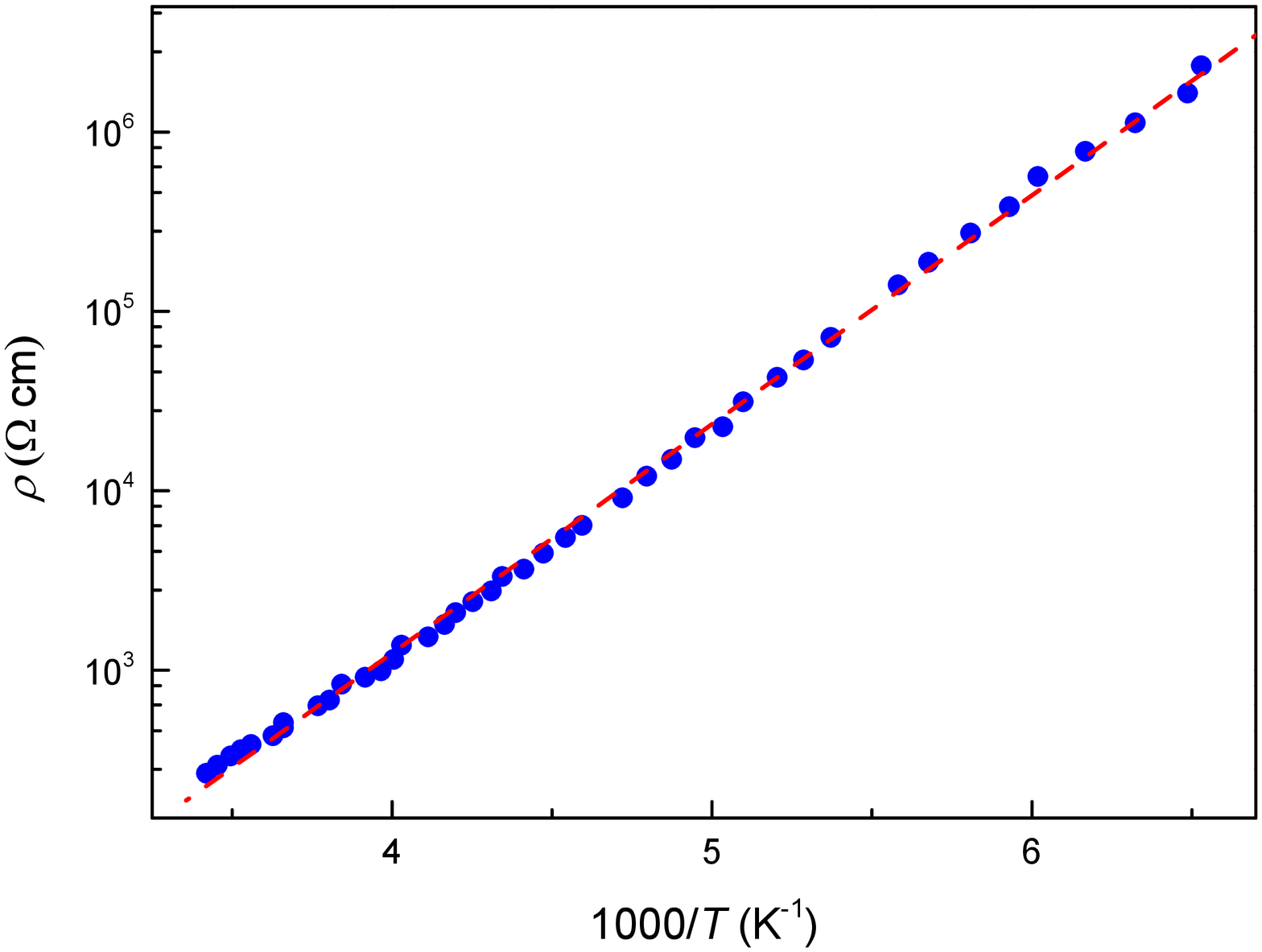}
\caption{\label{F2} Specific electrical resistivity  vs. temperature
for LaMnO$_3$. }
\end{figure}

Electrical resistivity has been measured for N1 sample by a
four-point probe method as well. In this case the $c$ axis of the
LMO single crystal is perpendicular to the current. Results are
presented in Fig.~3. In this case the resistivity of the N1 sample
shows the activation behavior at the whole experimental temperature
range. Results for N2 sample are shown in Fig.~4. Since the  $c$
axis of the LMO single crystal is perpendicular to the film plane,
the current is perpendicular to the $c$ axis. At high temperatures
the electrical resistivity of the N2 sample demonstrates activation
behavior, whereas at the temperature below ~160\,K it has
metallic-like features. At the same time, the total electrical
resistivity of the N2 sample is few times less than $R_0$ at the
whole experimental temperature range ($R_0$ is resistivity of the
sample without film, it was obtained by data from Fig.\,2 and
plotted in Fig.\,4). A sharp drop of the resistivity at low
temperature may be an evidence for 2DEG formation.
\begin{figure}
 \includegraphics[angle=-90,width=11.05cm]{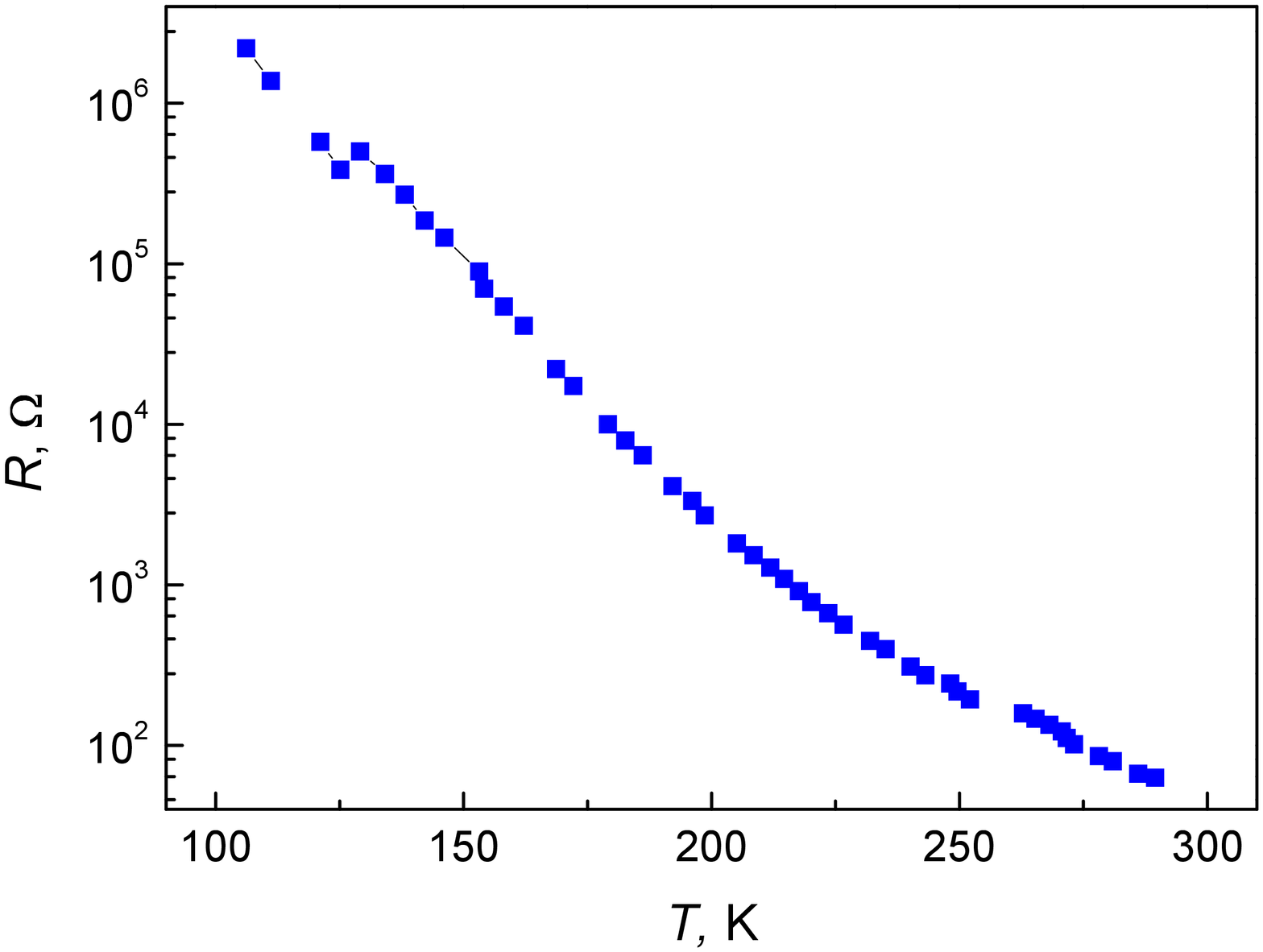}
\caption{\label{F3} Electrical resistivity vs. temperature for N1
sample (for a Ba$_{0.8}$Sr$_{0.2}$TiO$_3$/LaMnO$_3$
heterostructure). The thickness of a Ba$_{0.8}$Sr$_{0.2}$TiO$_3$
film is equal to 350 nm. }
\end{figure}
\begin{figure}
 \includegraphics[angle=-90,width=11.05cm]{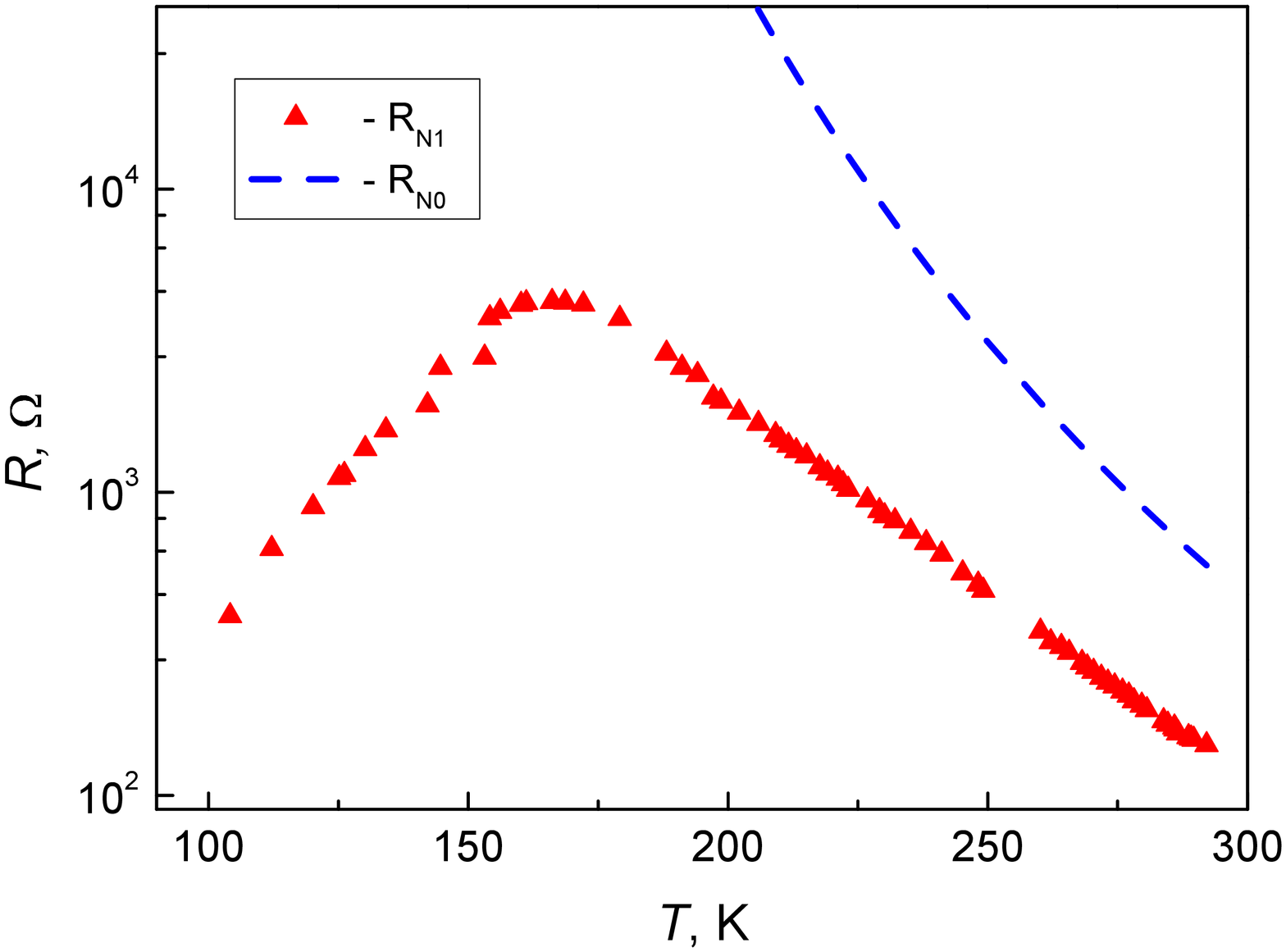}
\caption{\label{F4} Electrical resistivity vs. temperature for N2
sample (for a Ba$_{0.8}$Sr$_{0.2}$TiO$_3$/LaMnO$_3$
heterostructure). The thickness of the Ba$_{0.8}$Sr$_{0.2}$TiO$_3$
film amounts to 350 nm. Dashed line corresponds to the similar
LaMnO$_3$ sample without film.}
\end{figure}
Let us analyze the obtained results and perform some estimations of
the parameters of the emerging state. Suppose that the electrical
current at the whole measuring range flows over three layers with
different transport properties: (1) the first is the LMO single
crystal with the resistivity $R_0$ (is shown by dashed line in
Fig.\,4); (2) the second is the ferroelectric layer with a very high
resistivity (the current is almost zero) and (3) the last one is a
layer with a high charge carriers density inside the LMO slab near
the interface with a thickness $\Delta$ and resistivity $R_S$. Then,
$R_S$ can be calculated by formula {$R_S=R_0\,R/(R_0-R$)}, where $R$
is a measured resistivity of the sample. Near the temperature of the
liquid nitrogen $R_S \approx R\approx 200\, \Omega$. Taking into
account that:
\begin{equation}
R_S = \frac{\rho_m \cdot l }{d \cdot\Delta}.
\end{equation}
\noindent
Here $\rho_m$ is the specific  resistivity of the LMO in the
metallic state, $ l$ is the length of the area where conductivity
has been measured (distance between the potential contacts
$l=0.14\,$cm), $d $ is a transverse dimension of the measuring area
($d=0.33\,$cm). For a surface concentration of charge carriers
$n_s=n_v\cdot \Delta$, where $n_v$ is a volume charge carriers
concentration. Taking into account (1), $n_S$ can be rewritten as:
\begin{equation}
n_s = \frac{\rho_m \cdot  n_v  \cdot l}{R_S \cdot d}.
\end{equation}
\noindent If we take the number of the charge carriers (the doping
level) per unit cell of the order of  $x_0=0.175\div 0.3$, which is
typical for metallic conductivity of the lanthanum manganite at low
temperatures and write the volume charge carriers concentration as
$n_v=x_0\times(abc)^{-1}$ ($a,\,b,\,c\,$ are lattice parameters for
the LMO unit cell), then using known values for a manganite's
resistivity $\rho_m=2\times10^{-4}\div 6\times10^{-5}\,$\,Ohm$ \cdot
$cm near the liquid nitrogen temperature for compounds with similar
charge carriers density~\cite{S23}, we can obtain the surface charge
carriers concentration: $n_s=1.65\times10^{14}\div
3.03\times10^{14}$\,cm$^{-2}$. This gives the following estimation
of the layer thickness, where the high mobility electron gas forms:
$\Delta = n_s\times n_v^{-1}$, $\Delta \approx(1.65\div
5.51)\times\,c=1.265\div 4.216$\,nm ($c$=7.665\,\AA\, is LMO unit
cell size in the $c$-axis direction).

In conclusion, it should be noted that the regimes of deposition of
the BSTO films gives as a possibility to avoid the doping of
substrate by the elements from the film~\cite{S21, S22}. According
to this one can be sure that all observations are not due to
implantation of some elements (like Sr)  into wear-surface of the
single crystalline LMO. The fact that high mobility electron gas is
observed only for one LMO orientation (sample N2) and is not for
another orientation (sample N1) also indirectly confirms the
above-mentioned fact. Magnetic properties of the BSTO/LMO
heterostructure, effect of magnetic field on transport properties of
the 2DEG in this heterostructure, as well as ferroelectric layer
thickness effect on 2DEG properties  will be presented in next
publications.

Hence in our paper we have presented the calculations of the
structural and electric properties of the
ferroelectric/antiferromagnet (BaTiO$_3$/LaMnO$_3$) heterostructure.
The role of the structural reconstruction in formation of metallic
state at the interface is revealed. Electrical resistivity has been
measured for single crystalline antiferromagnetic LaMnO$_3$ samples.
After that the epitaxial ferroelectric Ba$_{0.8}$Sr$_{0.2}$TiO$_3$
film  was deposited on the LMO sample using the magnetron sputtering
technique. The measurements demonstrated that the resistivity of the
single crystalline antiferromagnetic LaMnO$_3$ sample with deposited
film of Ba$_{0.8}$Sr$_{0.2}$TiO$_3$ decreases strongly, and below
the temperature of 160\,K passes to a metallic-like behaviour. The
latter is observed only when the polarisation axis of the
ferroelectric film is perpendicular to the single crystal surface
and the film plane is perpendicular to the $\it c$ axis of the
LaMnO$_3$ single crystal. In this case, the substrate has a
compressive effect on the film.

Authors thank Yu.\,I.~Golovko for the assistance in the samples
characterisation, and S.\,А.~Migachev in the modification of the
single crystal surface before the film deposition. This study was
supported by the Supercomputing Center of Lomonosov Moscow State
University. The authors from Kazan Federal University acknowledge
partial support by the Program of Competitive Growth of Kazan
Federal University.

\end{document}